\documentclass[12pt]{article}
\usepackage[english]{babel}
\usepackage{graphicx,epsf}
\usepackage{amsmath}

\begin{document}

\renewcommand{\theequation}{\arabic{section}.\arabic{equation}}
\newcommand{\be}{\begin{eqnarray}}
\newcommand{\ee}{\end{eqnarray}}
\newcommand{\ben}{\begin{eqnarray*}}
\newcommand{\een}{\end{eqnarray*}}
\newcommand{\la}{\langle}
\newcommand{\ra}{\rangle}
\def\Sp{\mathop{\rm Sp}\nolimits}

\begin{center}
{\bf THEORY OF RADIATION AND ABSORPTION \\ OF DEFORMED FIELD QUANTA\footnote{This article has been written for the special issue
of the {\it Condensed Matter Physics} dedicated to the famous
Ukrainian theoretical physicist Professor Igor Stasyuk. Professor
Stasyuk exemplifies a highly erudite, passionate physicist who
never leaves any un-clarified aspects in physical and mathematical
mechanisms of models he proposes for the explanation of physical
phenomena. I greatly appreciate scientific (and beyond scientific)
discussions during the many years of our acquaintance which
started when we were students. These numerous discussions have
always been for me warm intellectual ``festivals'', and will
always stay as such in my heart.}}
\\ I.~O.~Vakarchuk\\
{\em Department for Theoretical Physics, Ivan Franko Lviv National
University, Drahomanov Str., 12, Lviv, UA--79005, Ukraine}
\end{center}

\smallskip
{\small We calculate the intensity of spontaneous radiation of a
system of non-linear quantum field, where the non-linearity is due
to deformations of the Poison brackets of the generalized
coordinates and momenta.}



\section{Introduction}
Starting from the seminal works by \cite{Sny1} and \cite{Sny2},
quantum systems with deformed Poison brackets have attracted much
attention in various fields of theoretical physics. Of a special
interest is the so called space with the ``minimal length,'' which
means that the deformation of the Poison bracket is quadratic in
momentum for both coordinates and momenta. This has first been
introduces in Refs.~\cite{kempf1,kempf2}. Such a deformation leads
to a non-zero minimal root-mean-square value of the coordinate.
There is an increasing number of papers on the subject, and the
literature review would require a separate article.

The deformation with the minimal length can in principle be
extended to an arbitrary space of generalized coordinates and
momenta which are natural for the description of various models.
In particular, one can study the electromagnetic field represented
as a set of oscillators with the deformed commutator relations for
coordinates and momenta, which generates the minimal length.
Obviously, in this case we have the deformation of the field
itself rather then the deformation of the real space. This model
of the electromagnetic field has been considered in
Refs.~\cite{camacho1,camacho2}. The Casimir effect in such a
deformed field has been studied in Ref.~\cite{vakarchuk}, where
the Casimir energy was calculated for the one-, two, and
three-dimensional cases. It has been shown that the deformation
suppresses the interaction between the confining boundaries.

In the present work, we study the interaction of the deformed
electromagnetic field with atomic systems in the coordinate
undeformed space. We shall not discuss the properties of the field
equations (which are non-linear) or problems related to their
Lorentz and gauge invariance which will be considered in a
separate paper, we shall rather propose a model of interaction of
such fields with atomic systems. We shall also calculate the
intensity of spontaneous radiation as a function of the
deformation parameter.

\section{Basic equations}

We study the interaction of an atomic system with the
electromagnetic field, which Hamiltonian after the decomposition
into a set of oscillators is given by:
\be
\hat H=\sum_{\bf k}\sum_\alpha\left(
    \frac{\hat P_{{\bf k},\alpha}^2}{2} +
    \frac{\omega_k^2 \hat Q_{{\bf k},\alpha}^2}{2}\right),
\label{v2.1} \ee where ${\bf k}$ is the wave-vector, $\alpha$
denotes the polarization, $\omega_k=ck$ is the frequency (where
$c$ is the speed of light in the vacuum). $\hat Q_{{\bf
k},\alpha}$, $\hat P_{{\bf k},\alpha}$ are the generalized
coordinate and momenta operators with the deformed Poison
brackets:
\be
\hat Q_{{\bf k},\alpha} \hat P_{{\bf k},\alpha}- \hat P_{{\bf
k},\alpha}\hat Q_{{\bf k},\alpha}= i\hbar \Big(1+\beta \hat
P_{{\bf k},\alpha}^2\Big), \label{1.2} \ee where $\beta \geq 0$ is
the deformation parameter, and all remaining commutators are equal
zero. As it is well-known, such commutation relations lead to the
``minimal length'' in the coordinate space, $\sqrt{\la\hat Q_{{\bf
k},\alpha}^2\ra}=\hbar\sqrt \beta$ \cite{kempf1,kempf2}. Because
in this case we do not deform the ordinary ``physical'' space (in
which our field lives) but rather the commutation relations of the
dynamical variables (fields), we have the deformation of the field
itself rather then the space-deformation.

Let us introduce new operators $\hat q_{{\bf k},\alpha}$ and $\hat
p_{{\bf k},\alpha}$ as follows:
\be
\label{eq:new_operators} \hat q_{{\bf k},\alpha} = \hat Q_{{\bf
k},\alpha},\ \ \ \ \ \hat P_{{\bf k},\alpha} = \frac{1}{\sqrt
\beta}\tan (\hat p_{{\bf k},\alpha}\sqrt \beta). \ee It is easy to
show that they are canonically conjugate operators:
\be
\hat q_{{\bf k},\alpha}\hat p_{{\bf k},\alpha}-\hat p_{{\bf
k},\alpha}\hat q_{{\bf k},\alpha}=i\hbar. \ee The Hamiltonian
(\ref{v2.1}) becomes
\be
\hat H =\sum\limits_{\bf k}\sum\limits_\alpha \left(
\frac{\omega_k^2 \hat q_{{\bf k},\alpha}^2}{2}
    + \frac{\tan^2(\hat p_{{\bf k},\alpha}\sqrt \beta)}{2\beta}\right).
\label{1.5} \ee

The equation of motion of such a field have been studied in
Ref.~\cite{camacho1,camacho2}, where the Hamiltonian was presented
as an expansion in ordinary annihilation and creation operators.
The field equations are non-linear but can be treated
perturbatively.

As it is usual for oscillatory systems, we present the magnetic
vector potential ${\bf A}$ as follows
\be
{\bf A} = \sqrt{\frac{4\pi c^2}{V}}\sum_{\bf k}
    \sum_\alpha {\bf e}_{{\bf k},\alpha}
      \left[\frac{1}{2} \left(\hat Q_{{\bf k},\alpha} - \frac{\hat P_{{\bf k},\alpha}}{i\omega_k}\right) e^{i{\bf kr}}
      + \frac{1}{2} \left(\hat Q_{{\bf k},\alpha} + \frac{\hat P_{{\bf k},\alpha}}{i\omega_k}\right)
                    e^{-i{\bf kr}}\right].
\ee In terms of the new, canonically conjugate operators
(\ref{eq:new_operators}) we have:
\be
{\bf A} = \sqrt{\frac{4\pi c^2}{V}}\sum_{\bf k}
    \sum_\alpha {\bf e}_{{\bf k},\alpha}
    \left[\hat q_{{\bf k},\alpha}\cos({\bf kr})
    - \frac{\tan (\hat p_{{\bf k},\alpha}\sqrt\beta)}{\omega_k\sqrt\beta}
    \sin({\bf kr})\right].
\ee

As usual, the operator of the interaction of the field with an
atom is:
\be
\hat V = -\frac{e}{mc}({\bf A}\hat{\bf p}) + \frac{e^2}{2mc^2}{\bf
A}^2, \label{v2.8} \ee where $\hat {\bf p}$ is the momentum
operator (of the electron in an atom), $e=-|e|$ is the electron
charge and $m$ the mass.

Therefore, we propose the following model of the interaction of
the deformed field with an atomic system: we assume that all the
expressions of the ordinary theory of the electromagnetic field
are valid, except that there are deformed commutation relations
for the operators $\hat Q_{{\bf k},\alpha}$ and $\hat P_{{\bf
k},\alpha}$. Such a non-linear field is a simple model which
allows us to study the effects of the deformation of the Poison
brackets on the properties of physical systems.

\section{Wave function and energy levels of the deformed field}

The eigenvalue problem for the Hamiltonian (\ref{1.5}) has an
analytic solution. In momentum representation the wave function
is:
\be
\psi_{\ldots,N_{{\bf k},\alpha},\ldots}(\ldots,p_{{\bf
k},\alpha},\ldots)=\prod_{\bf k}\prod_\alpha \psi_{N_{{\bf
k},\alpha}}(p_{{\bf k},\alpha}), \ee where $\psi_{N_{{\bf
k},\alpha}}(p_{{\bf k},\alpha})$ is the eigenfunction for the
$({\bf k},\alpha)$ mode of the Hamiltonian
\be
\hat H = \frac{\omega^2\hat q^2}{2} + \frac{\tan^2(\hat
p\sqrt\beta)}{2\beta}, \ee which can be calculated using the
standard methods, for instance, the factorization method
\cite{cooper}. For brevity, we drop indexes ${\bf k},\alpha$ from
operators $\hat q_{{\bf k},\alpha}$, $\hat p_{{\bf k},\alpha}$,
frequency $\omega_{k}$, quantum number $N_{{\bf
k},\alpha}=0,1,2,\ldots$ and other quantities:
\be
p_{{\bf k},\alpha}\to p,\quad \hat q_{{\bf k},\alpha} =
i\hbar{\partial\over \partial p_{{\bf k},\alpha}}\to i\hbar
{\partial\over \partial p},\quad \omega_k \to \omega, \quad
N_{{\bf k},\alpha}\to n=0,1,2,\cdots . \ee

We have therefore for the wave function in the momentum
representation:
\be
&&\psi_n(p)=\beta^{1/4}\sqrt{\Gamma(\alpha+n+1)\Gamma(n+2\alpha)\over
n!\Gamma(1/2)\Gamma\left(\alpha+n+{1/2}\right)\Gamma(2n+2\alpha)}\nonumber\\
&&\times \left(-{d\over d\bar p}+\alpha\tan{\bar
p}\right)\ldots\left(-{d\over d\bar p}+(\alpha+n-1)\tan{\bar
p}\right)\cos^{\alpha+n}\bar p, \label{v2.12} \ee where the
dimensionless momentum $\bar p=p\sqrt \beta$, with the domain of
definition of the momentum $p$ being
\be
-\pi/2\sqrt\beta\le p\le \pi/2\sqrt\beta. \ee The quantum number
is $n=0,1,2,\ldots$, and
\be
\alpha={1\over 2}\left[1+{2\over \beta\hbar
\omega}\sqrt{1+\left({\beta\hbar \omega\over 2}\right)^2}\right].
\label{v14}
\ee

The wave-function (\ref{v2.12}) is normalized as usually,
\be
\int\limits_{-\pi/2\sqrt\beta}^{\pi/2\sqrt\beta}
\psi_{n'}(p)\psi_n(p)\,dp=\delta_{n'n}. \ee

A few first wave-functions read:
\be
&&\psi_0(p)=\beta^{1/4}\sqrt{\Gamma(\alpha+1)\over
\Gamma(1/2)\Gamma(\alpha+1/2)}\cos^\alpha\bar p,\nonumber\\
&&\psi_1(p)=\beta^{1/4}\sqrt{2\Gamma(\alpha+2)\over
\Gamma(1/2)\Gamma(\alpha+1/2)}\cos^\alpha \bar p\sin\bar
p,\label{v3.13}\\
&&\psi_2(p)=\beta^{1/4}\sqrt{2\Gamma(\alpha+3)\over
2(\alpha+1)\Gamma(1/2)\Gamma(\alpha+3/2)}\left[(2\alpha+1)-2(\alpha+1)\cos^2\bar
p\right] \cos^\alpha \bar p.\nonumber \ee

Let us calculate the wave-functions for $\beta=0$. In this case
there is no deformation and the Hamiltonian (\ref{1.5}) reduces to
the Hamiltonian of a harmonic oscillator with the mass equal to
unity. Because $\alpha\to 1/\beta\hbar\omega\to \infty$ for
$\beta\to 0$, we obtain
\be
\cos^\alpha(p\sqrt\beta)\mathop{=}\limits_{\beta\to 0}
\left(1-{p^2\beta\over
2}+\ldots\right)^{1/\beta\hbar\omega}\mathop{=}\limits_{\beta\to
0}
 e^{-p^2/2\hbar\omega}.
\ee Noticing that $\Gamma(z +a)\sim\sqrt{2\pi}e^{-z}z^{z+a-1/2}$
for $z\to \infty$, we obtain from Eq.~(\ref{v2.12}), as expected,
the wave-fucntion of the harmonic oscillator in the momentum
representation:
\be
\psi_n(p)={1\over (\pi\hbar\omega)^{1/4}} {1\over
\sqrt{n!2^n}}\left(-{d\over d\eta}+\eta\right)^ne^{-\eta^2/2}, \ee
where the dimensionless variable  $\eta=p/\sqrt{\hbar\omega}$,
with $-\infty<p<\infty$.

It is convenient to rewrite Eq.~(\ref{v2.12}) using the Gegenbauer
polynomials $C_n^\alpha(x)$ \cite{janke,gradshteyn}. To do this,
we change the variable $x=\sin\bar p$ to obtain, after simple
transformations,
\be
\bar\psi_n(x)=\sqrt{n!(\alpha+n)\over 2\pi
\Gamma(2\alpha+n)}2^\alpha
\Gamma(\alpha)(1-x^2)^{\alpha/2-1/4}C_n^\alpha(x), \ee

\be
C_n^\alpha(x)&=&(-)^n{\sqrt\pi \Gamma(2\alpha+n)\over
n!2^{2\alpha+n-1}\Gamma(\alpha)\Gamma(\alpha+n+1/2)}(1-x^2)^{1/2-\alpha}\left({d\over
dx}\right)^n\nonumber\\
&\times& \left[(1-x^2)^{n+\alpha-1/2}\right]. \ee
The wave-functions have the following normalization
\be
\int\limits_{-1}^1 \bar\psi_{n'}(x)\bar\psi_n(x)\,
dx=\delta_{n'n}. \ee Our change of the variable is not a unitary
transformation; the Jacobian of the transformation is $1/\cos \bar
p=(1-x^2)^{-1/2}$. Hence, for instance, the coordinate operator in
the coordinate representation is
\be
\hat q={i\hbar \sqrt \beta} (1-x^2)^{1/4}{d\over dx}(1-x^2)^{1/4}.
\ee

The energy levels of the harmonic oscillator with the commutation
relations (\ref{1.2}) are eigenvalues of the Hamiltonian
(\ref{1.5}). They are well-known\cite{kempf2,vak}. Then the energy
levels of the deformed field read:
\be
\label{eq:energy_deformed} E_{\ldots,N_{{\bf
k},\alpha},\ldots}&=&\sum_{\bf k}\sum_\alpha \hbar
\omega_k\Bigg[\left(N_{{\bf k},\alpha}+{1\over 2}\right)
\sqrt{1+\left({\beta \hbar \omega_k\over 2}\right)^2}\nonumber
\\
&+&{\beta \hbar \omega_k\over 2} \left(N_{{\bf
k},\alpha}^2+N_{{\bf k},\alpha}+{1\over 2}\right)\Bigg], \ee where
the quantum numbers $N_{{\bf k},\alpha}=0,1,2,\ldots\;$. We note
that the deformation parameter may depend on ${\bf k}$ and
$\alpha$. In our work, however, we restrict our considerations to
the case $\beta=\mathrm{const}$.

\section {Spontaneous radiation}

Let us assume that an atom is in the excited state $|2\ra$ (with
the energy $E_2$) while the field is in the ground state
$|\ldots,0,\ldots\ra$. As a result of the interaction of the atom
with the field, the atom jumps to a level with the lower energy
$E_1$ and radiates a light quantum with the energy $\hbar
\omega=E_2-E_1$. The field, therefore, jumps to the state with one
phonon, i.e., $|\ldots,0,N_{{\bf k},\alpha}=1,0\ldots\ra$. The
transition probability rate of a system ``field-plus-atom'' to
jump from the initial state $|i\ra=|2\ra|\ldots,0,\ldots\ra$ to
the final state $|f\ra=|1\ra|\ldots,0,N_{{\bf
k},\alpha},0,\ldots\ra$  is given by: \be w_{k\to f}={2\pi\over
\hbar}\left({e\over mc}\right)^2\left|\la f|{\bf A}\hat {\bf
p}|i\ra\right|^2\delta (E_2-E_1-\hbar \Omega_k), \label{v4.20} \ee
where the energy of the field quantum, according to
Eq.~(\ref{eq:energy_deformed}), is
\be
\hbar \Omega_k&=&E_{\ldots,0,N_{{\bf
k},\alpha}=1,0,\ldots}-E_{\ldots,0,\ldots},\nonumber\\ \nonumber
\\ \Omega_k&=&\omega_k\left[\sqrt{1+\left(\beta{\hbar
\omega_k\over 2}\right)^2}-\beta\hbar \omega_k\right].
\label{v4.21} \ee We take into account the first term of the
interaction operator (\ref{v2.8}), which is linear in the vector
potential.

The matrix element $\la f|{\bf A\hat p}|i\ra$ in Eq.~(\ref{v4.20})
is \ben \la f|{\bf A\hat p}|i\ra=\sqrt{4\pi c^2\over V}\left(\la
N_{{\bf k},\alpha}=1|\hat q_{{\bf k},\alpha}|0\ra p_{12}^c-
{1\over \omega_k\sqrt\beta}\la N_{{\bf k},\alpha}=1\left|\tan
(p_{{\bf k},\alpha}\sqrt\beta)\right|0\ra p_{12}^s\right), \een

\be p_{12}^c=\la 1|\cos({\bf kr}){\bf e}_{{\bf k},\alpha}\hat {\bf
p}|2\ra,\quad p_{12}^s=\la 1|\sin({\bf kr}){\bf e}_{{\bf
k},\alpha}\hat {\bf p}|2\ra. \label{v4.22} \ee

Using the wave functions (\ref{v3.13}), we get for the matrix
elements of the field operators:
\be
\la N_{{\bf k},\alpha}=1|\hat q_{{\bf k},\alpha}|0\ra=-i\hbar
{\sqrt{2\beta(\alpha_k+1)}\over
2\alpha_k+1}\left[{\Gamma(\alpha_k+1)\over \Gamma(\alpha_k+1/2)}
\right]^2, \nonumber\\ \label{v4.23}
\\ \nonumber
\la N_{{\bf k},\alpha}=1\left|\tan (p_{{\bf
k},\alpha}\sqrt\beta)\right|0\ra ={\sqrt{2(\alpha_k+1)}\over
\alpha_k(2\alpha_k+1)}\left[{\Gamma(\alpha_k+1)\over
\Gamma(\alpha_k+1/2)} \right]^2, \ee here $\alpha_k=\alpha$,
which is defined by (\ref{v14}) at $\omega\to \omega_k$.

The intensity of spontaneous radiation $I_{{\bf k},\alpha}$ is
defined in a usual way as the amount of energy with the given
polarization $\alpha$ radiated by an atom in unit time on the
resonance frequency $\omega=(E_2-E_1)/\hbar$ per space angle,
i.e.,
\begin{align}
\label{eq:intensity:def} I_{{\bf k},\alpha} =
    {V\over (2\pi)^3}\int_0^\infty k^2\hbar \Omega_k w_{i\to f}\, dk.
\end{align}

By inserting Eqs.~(\ref{v4.20})--(\ref{v4.23}) into
Eq.~(\ref{eq:intensity:def}) we obtain, after integration: \ben
I_{{\bf k},\alpha}={e^2\hbar\over m^2c^3\pi}\left(
{2\beta(\alpha_k+1)\over
(2\alpha_k+1)^2}\left[{\Gamma(\alpha_k+1)\over
\Gamma(\alpha_k+1/2)} \right]^4\omega_k^2\Omega_k{d\omega_k\over
d\Omega_k}\left|p_{12}^c-{i\over \beta\alpha_k\hbar \omega_k}
p_{12}^s\right|^2\right)_{\Omega_k=\omega}. \een

Taking into account Eq.~(\ref{v4.21}), we finally get, after
simple transformations,
\be
I_{{\bf k},\alpha}={e^2\omega^2\over 2\pi m^2c^3} g(\bar\omega)
\left|p_{12}^c-ip_{12}^s{1\over \bar\alpha}\sqrt{2\bar
\alpha+1\over 4\bar\omega}\right|^2, \label{v4.24} \ee where
\be
g(\bar\omega)={8(\bar\alpha+1)(2\bar\alpha-1)\bar\omega^{1/2}\over
[2\bar\alpha+1+2\bar\omega(4\bar\alpha-1)](2\bar\alpha+1)^{5/2}}
\left[{\Gamma(\bar\alpha+1)\over \Gamma(\bar\alpha+1/2)}
\right]^4, \ee \ben \bar\alpha={3\bar \omega\over
1+4\bar\omega-\sqrt{1+8\bar\omega+4\bar\omega^2}}-{1\over 2}, \een
and the dimensionless deformation parameter is
\be
\bar \omega={\beta \hbar \omega\over 2}. \label{v4.26} \ee

In the limiting case of no deformation ($\beta=0$) we have
$g(0)=1$ (see Eq.~(\ref{v4.24})), and therefore one gets for the
intensity
\be
I_{{\bf k},\alpha}={e^2\omega^{2}\over 2\pi
m^2c^3}|p_{12}|^2,\quad p_{12}=\la 1|e^{i{\bf kr}}({\bf e}_{{\bf
k},\alpha} \hat {\bf p})|2\ra, \label{v4.27} \ee which is the
well-known expression for the intensity of radiation for the
non-deformed field.

In the opposite limit of large deformation parameter ($\bar
\omega\gg 1$), the term $\sim \sin({\bf kr})$ vanishes (see
Eq.~(\ref{v4.24})), and hence the quadrupole radiation vanishes
too. The reason is that the $\sin({\bf kr})$ term is the main
contribution to the quadrupole radiation in the long wavelength
limit. Finally we note that the function $g(\bar \omega)$ in the
case of considerable deformations ($\bar\alpha=1$) has the
following asymptotic form:
\be
\label{eq:g(w)} g(\bar\omega)\mathop{=}\limits_{\bar\omega\to
\infty}{128\over 27\pi^2}{1\over \sqrt {3\bar\omega}}. \ee

\section{Intensity of the dipole radiation}

Within the dipole-transition approximation, the wave-vector ${\bf
k}\to0$ ($\cos({\bf k r})\to 1$ and $\sin({\bf kr})\to 0$), hence
$p_{12}^c=({\bf e}_{{\bf k},\alpha} {\bf p}_{12})$ and
$p_{12}^s=0$ (see Eq.~(\ref{v4.22})). We therefore get:
\be
I_{{\bf k},\alpha}={e^2\omega^2\over 2\pi m^2c^3}| {\bf e}_{{\bf
k},\alpha} {\bf p}_{12}|^2g(\bar\omega), \label{v5.29} \ee
Equation (\ref{v5.29}) leads to Eq.~(\ref{v4.27}) for
$\bar\omega=0$ with ${\bf k}=0$, while for large values of the
deformation parameter the intensity of the dipole radiation is
(after making use of Eq.~(\ref{eq:g(w)}))
\be
I_{{\bf k},\alpha}={e^2\omega^{3/2}\over 2\pi m^2c^3}|{\bf
e}_{{\bf k},\alpha} {\bf p}_{12}|^2 \left({2\over
3}\right)^{7/2}{1\over \sqrt{\beta\hbar}}\left({4\over
\pi}\right)^2. \ee We therefor conclude that with increasing of
the deormation parameter the intensity decreases as
$\beta^{-1/2}$.

\begin{figure}[htb]
\centerline{\includegraphics[width=0.7\textwidth]{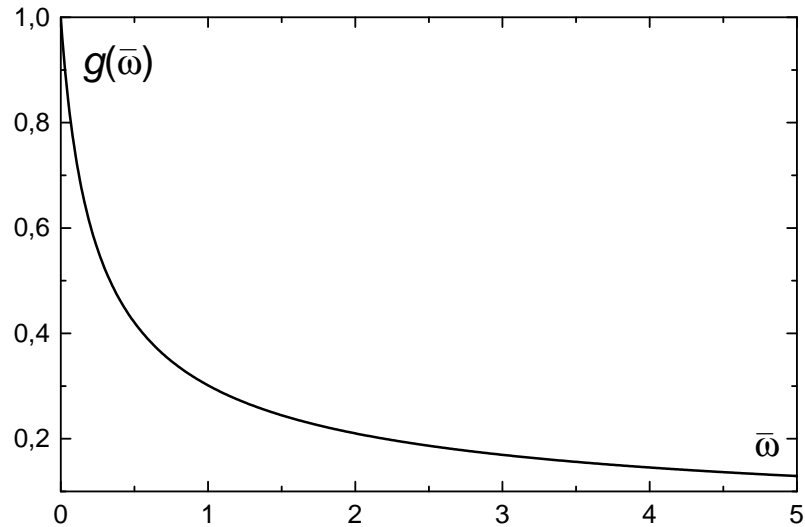}}
\caption{The intensity of the spontaneous dipole radiation of the
deformed field (see Eq.~(\ref{v5.29})), divided by the intensity
for the undeformed field, as a function of the dimensionless
deformation parameter~(\ref{v4.26}).}  \label{fig:intensity}
\end{figure}

The intensity of the dipole interaction as a function of the
dimensionless deformation parameter $\bar\omega$ is shown in
Fig.~\ref{fig:intensity}

\section{Conclusion}

We have considered a theory of spontaneous radiation taking into
account the creation of one photon. Because the field is
non-linear, the transitions between any two states are allowed,
not only between the two neighboring, as in the case of the
ordinary oscillator. In order to study such processes in the case
of more then one photon, but in the linear approximation (in the
vector potential ${\bf A}$ in the interaction operator
(\ref{v2.8})) and neglecting all the higher order terms in an
expression for the quantum transition rates, we take the following
matrix elements instead of (\ref{v4.23}):
\begin{align}
\label{eq:matrix_elements2} q_{n,n'} = {i\hbar\sqrt\beta}
\int_{-1}^1 \bar \psi_{n}(x)(1-x^2)^{1/4}\,{d\over dx}
    \left[\bar\psi_{n'}(x)(1-x^2)^{1/4}\right]\,dx,\\
\left(\tan\,\bar
p\right)_{n,n'}=\int_{-1}^1\bar\psi_{n}(x)(1-x^2)^{1/4}{x\over
\sqrt{1-x^2}}\bar\psi_{n'}(x)(1-x^2)^{1/4}\,dx.
\end{align}
It is obviously that these integrals are not equal zero only at
even $n+n'$.

In general case the integrals in Eq.~(\ref{eq:matrix_elements2})
cannot be expressed in terms of elementary functions. For some
particular values of $n$ and $n'$, however, the integrals can be
easily evaluated, as for instance for $n=1$ and $n'=0$ (see
Eq.~(\ref{v4.23})). Evidently, many-photon processes are absent
for $\beta = 0$ while their role increases with increasing the
deformation parameter $\beta$.

\section*{Acknowledgements}
I wish to thank Prof. I.~Stasyuk and Dr. V.~Tkachuk for
stimulating discussions, especially during the 12th {\it Christmas
Discussions} (2008) held at the Department of Theoretical Physics
of the Lviv National University.

\end{document}